\documentclass[10pt,aps,prd,nofootinbib,superscriptaddress,twocolumn,preprintnumbers,balancelastpage]{revtex4}
\usepackage{amssymb,amsmath,latexsym,graphics, graphicx,epsfig,multirow,comment,hyperref,appendix,feyn,slashed}

\newcommand {\be} {\begin {equation}}
\newcommand {\ee} {\end {equation}}

\newcommand {\bes} {\begin {equation*}}
\newcommand {\ees} {\end {equation*}}

\newcommand{\es}[2] {\begin{equation} \label{#1} \begin{split} #2 \end{split} \end{equation}}

\newcommand{\beq}{\begin{equation}}
\newcommand{\eeq}{\end{equation}}

\begin{document}

\title{Directional Antineutrino Detection}
\preprint{
MIT-CTP-4605}

\author{Benjamin R. Safdi}
\affiliation{Center for Theoretical Physics, Massachusetts Institute of Technology, Cambridge, MA 02139}
\email{bsafdi@mit.edu}

\author{Burkhant Suerfu}
\affiliation{Department of Physics, Princeton University, Princeton, NJ 08544}
\email{suerfu@princeton.edu}

\date{\today}

\begin{abstract}
	We propose the first truly directional antineutrino detector for antineutrinos above the hydrogen inverse beta decay (IBD) threshold, with potential applications including monitoring for nuclear nonproliferation, spatially mapping geo-neutrinos, characterizing the diffuse supernova neutrino background, and searching for new physics in the neutrino sector.
	The detector consists of adjacent and separated target and capture scintillator planes. IBD events take place in the target layers, which are thin enough to allow the neutrons to escape without scattering elastically. The neutrons are detected in the thicker, boron-loaded capture layers.
	The location of the IBD event and the momentum of the positron are determined by tracking the positron's trajectory through the detector.
	Our design is a straightforward modification of existing antineutrino detectors; a prototype could be built with existing technology. 

\end{abstract}
\maketitle

We present the first realistic proposal for directional antineutrino detection, through a design we call SANTA (Segmented AntiNeutrino Tomography Apparatus).
Such a detector would have immediate applications monitoring nuclear reactors for 
 nonproliferation (see, for example,~\cite{Bernstein:2001cz,Nieto:2003wd,Bernstein:2009ab,Christensen:2013eza}) and imaging the cores of nuclear reactors and 
 radioactive waste.  
Moreover, the directionality significantly cuts down on background compared to non-directional detectors.
The reduced-background properties of directional detectors make them ideal detectors for short baseline neutrino experiments searching for new physics in the neutrino sector, such as IsoDAR/DAEdALUS~\cite{Aberle:2013ssa,Bungau:2012ys}.

A large-volume SANTA, with hundreds of tones of target mass, would be capable of spatially mapping geo-neutrinos~\cite{Fiorentini:2007te} and thus constructing a map of radioactive material inside the Earth. Geo-neutrinos have been detected at the KamLAND~\cite{Araki:2005qa} and Borexino experiments~\cite{Bellini:2010hy}, but these experiments lack directionality.
	Other applications of such a detector to fundamental physics include searching for solar antineutrinos that could indicate neutrino electromagnetic interactions~\cite{PhysRevD.24.1883,PhysRevD.25.283} and characterizing the predicted diffuse supernova neutrino background~\cite{zeld,0034-4885-28-1-312,okh} (see~\cite{2010ARNPS..60..439B} for a recent review).

	Low-energy antineutrinos, with energies $\sim$2--10 MeV, are typically detected by inverse beta decay (IBD). The antineutrino scatters inelastically with a proton into a neutron and a positron. The positron quickly loses energy and annihilates with an electron. The neutron diffuses for a longer time before it reaches thermal speeds and is captured.
 
 Current detectors cannot determine the antineutrino's direction on an event-by-event basis because of neutron diffusion. The neutron recoils in approximately the direction of the antineutrino's velocity. However, by the time it is captured the neutron has little preference to end up in the direction it was originally traveling.
Still, some detectors have been able to use statistical methods to extract directional information about the distribution of antineutrinos, including Gosgen~\cite{Zacek:1986cu}, Bugey~\cite{Declais:1994su}, Palo Verde~\cite{Boehm:2001ik}, and  CHOOZ~\cite{Apollonio:2002gd}. 
	For example, with $\sim$$2500$ total IBD events, the CHOOZ experiment was able to determine the direction of the nuclear power plant where the antineutrinos were produced to within $\sim$$18^0$ at 68\% C.L.~\cite{Apollonio:2002gd}. The CHOOZ experiment used a 0.09\% Gd-loaded liquid scintillator target to minimize the neutron diffusion length. Recently,~\cite{Tanaka:2014fk} studied the advantages of using $^6$Li-loaded scintillators to increase position resolution and directional sensitivity; they concluded that small improvements in the angular resolution compared to CHOOZ may be possible in the future. The mini-Time-Cube project~\cite{mini} plans to use boron-loaded plastic scintillators to improve their directional sensitivity.

\begin{figure}[htb]
	\leavevmode
	\begin{center}$
	\begin{array}{c}
	\scalebox{.4}{\includegraphics{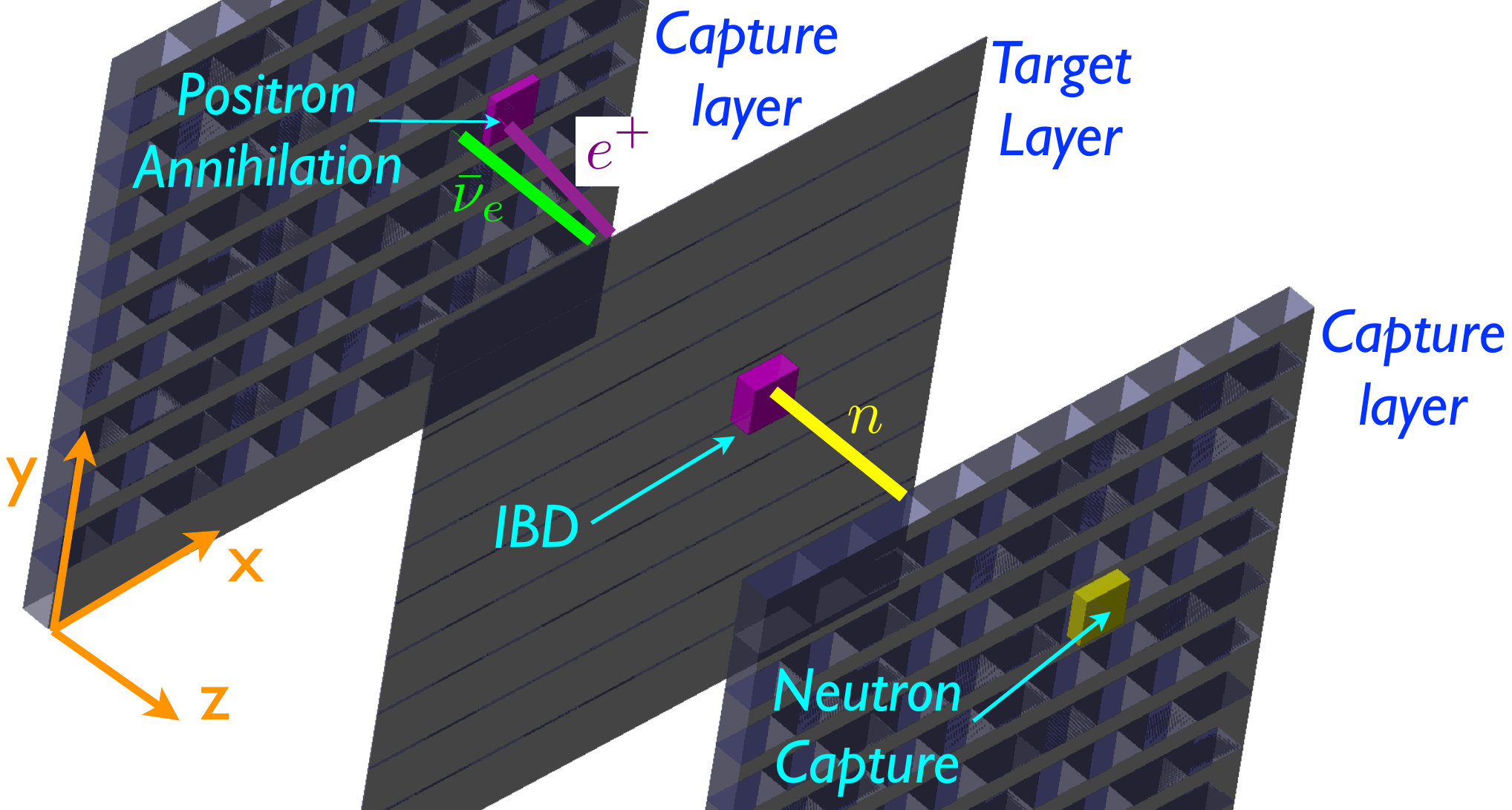}} 
	\end{array}$
	\end{center}
	\vspace{-.50cm}
	\caption{The detector consists of alternating layers of plastic scintillator, with the capture layers loaded with boron.
	 IBD events take place in the thin target layers, and the positron subsequently deposits energy (purple boxes) within the target layer and travels to the adjacent, thick capture layer, where it annihilates.  
The neutron propagates freely to the capture layer, where it diffuses and is captured on $^{10}$B, depositing energy (yellow box), with a delayed coincidence from the positron annihilation.}      
	\vspace{-0.15in}
	\label{Fig: 1}
\end{figure}

	{\it Detector concept}---We present a simple detector concept that circumvents the neutron diffusion limitation of previous detectors, which we refer to as monolithic detectors. 
	The idea is to make the target, where IBD events occur, a thin enough sheet of scintillator so that most neutrons escape without scattering elastically, therefore preserving the directional information.
	The neutrons then travel through free space to adjacent capture layers, where they diffuse and are captured (see Fig.~\ref{Fig: 1}).
The IBD location and the neutron capture location can be used to deduce the direction of the neutron's momentum ${\bf p_n}$. In this Letter, we take the region between layers to be vacuum for simplicity. However, this region may be any low-density medium, such as air, so long as the probability of neutron elastic scattering is small.  Charged-particle tracking may also be introduced between layers.

The IBD event location is determined from the positron, which deposits energy within the target layer through ionization, Bhabha scattering, and Bremsstrahlung.
The positrons may either annihilate within the target layer or escape, traverse between layers, and then lose energy and annihilate in one of the capture layers.
 	The annihilation results in
	 two back-to-back $\sim$$0.5$ MeV $\gamma$'s.
	 Within a few ns of the IBD event, there may be multiple coincident signals from the positron alone.
	  The positron's energy $E_{e^+}$ is measured from the total energy deposited in the detector in this short time. When the positron escapes the target layer, the direction of the positron's momentum ${\bf p_{e^+}}$ may be reconstructed from the spatial and temporal distribution of deposited energy.  Charged-particle tracking between layers may also be used to reconstruct ${\bf p_{e^+}}$.  It may also be possible to determine ${\bf p_{e^+}}$ within the target layer itself by drifting the secondary ions produced by the positron towards the target-layer sides and measuring the distribution of arrival times and locations, for example~\cite{Dawson:2014lga}.
	In the remainder of this Letter, we use Monte-Carlo simulations in GEANT4~\cite{2003NIMPA.506..250A} to demonstrate the directional capability for a specific SANTA configuration.

{\it Detector simulations}---We take the target and capture layers to be plastic scintillators, with the capture layers loaded with 5\% natural B by weight, which is commercially available~\cite{eljen}. The $^{10}$B is introduced for its high neutron-capture cross section. Moreover, neutron capture on $^{10}$B results in an $\alpha$, $\gamma$, and $^7$Li, with a Q value $\sim$2.78 MeV.  The majority of this energy is deposited within a very short distance in the scintillator, which helps identify the neutron capture.

 \begin{figure*}[htb]
	\leavevmode
	\begin{center}$
		\begin{array}{c}
		\scalebox{.54}{\includegraphics{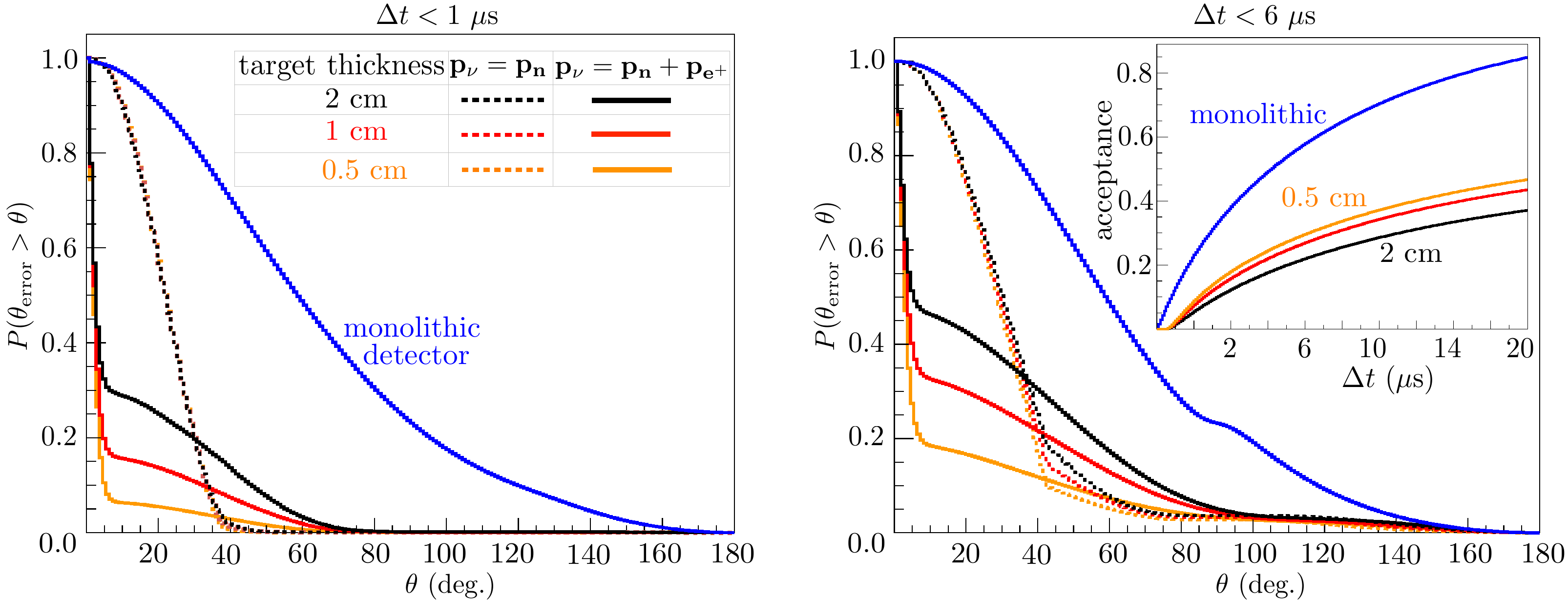}} 
		 \end{array}$
	\end{center}
	\vspace{-.50cm}
	\caption{We calculate $P(\theta_\text{error} > \theta)$ for $4$ MeV antineutrinos incident normal to the target plane for $0.5$, $1$, and $2$ cm thick target layers, with $\theta_\text{error}$ the angular error in the reconstruction of ${\bf \hat p_\nu}$: $\cos \theta_\text{error}=\bf \hat p_\nu \cdot \bf \hat p_\nu^\text{Rec.}$.  The antineutrino momentum is reconstructed using two methods.  The first method only uses the neutron's reconstructed momentum, ${\bf \hat p_\nu^\text{Rec.}} = {\bf \hat p_n}$, while the second method uses the reconstructed neutron momentum and the positron momentum, which we assume is reconstructed exactly: ${\bf p_\nu^\text{Rec.}} = {\bf p_n} + {\bf p_{e^+}}$.  A key tool for improving the angular resolution is timing.  The left panel imposes a $< 1$ $\mu$s timing cut between the positron annihilation and the neutron capture, while the right panel uses a $<6$ $\mu$s timing cut.  The stricter timing cuts, however, result in a reduced fraction of events that are accepted by the analysis, as shown in the inset plot on the right panel.  For reference, we also show the angular resolution and acceptance rate for a boron-loaded monolithic detector assuming perfect reconstruction of the neutron-capture location and the IBD event location.}
	\vspace{-0.15in}  
	\label{Fig: 3}
\end{figure*}   

	In practice, each layer may consist of stacks of long, thin scintillator bars, similar to the PANDA antineutrino experiment~\cite{2012NIMPA.690...41K}, the PROSPECT experiment~\cite{Djurcic:2013oaa}, and the DANSSino experiment~\cite{Belov:2013qwa}.  Position resolution along the directions of the scintillator bars may be achieved using timing and by comparing the luminosity at the two ends.
We do not model the position resolution within the scintillator sheets in our simulations, as this depends heavily on the specific experimental configuration.
	
	The target layer should be thin enough for most neutrons to escape without elastically scattering off hydrogen or carbon; this corresponds to a target-layer thickness $\sim$1 cm in our material.  We illustrate target-layer thicknesses of $0.5$, $1$, and $2$ cm.
	Most neutrons are captured on $^{10}$B within a few cm in the boron-loaded plastic scintillator. For definiteness, we take the capture layers to be $6$ cm thick.  With this thickness, only $\sim$5\% of 50 keV neutrons incident normal to the capture layer pass through the layer without capture.\footnote{Neutrons recoiling from reactor-energy antineutrinos have kinetic energies $\sim$1--100 keV.} We take the layers to be separated by $1$ meter, as this is much longer than the thicknesses of each individual layer. Better angular resolution may be achieved by using a longer separation.
  
	A finite energy threshold in the scintillator layers would introduce a source of background, where IBD occurs in the capture layer, but the positron escapes, depositing less energy than the threshold, and then travels through the adjacent target layer. For typical energy thresholds $\sim$200 keV and below and a target-layer thickness $\sim$1 cm, this source of background is negligible, since a positron deposits $\sim$MeV of energy through ionization per cm in plastic. Charged-particle tracking between layers would eliminate this background completely. Another way of eliminating this background would be to use a neutron detector without hydrogen, such as a $^3$He neutron detector. 

The reconstruction of ${\bf p_{e^+}}$ is straightforward once the positron has left the target layer.
However, hard scattering within the target layer may deflect the positron before it leaves that layer. Our ability to account for hard scattering within the target layer is sensitive to the specific detector design and energy thresholds. To keep our analysis general, we reconstruct the antineutrino's momentum in two ways. First, we use the neutron's direction alone and equate the unit vectors \mbox{${\bf \hat p_\nu} \approx {\bf \hat p_n}$}, where $ {\bf \hat p_n}$ points in the direction of the neutron's reconstructed momentum. 
Second, we assume that we may exactly reconstruct
${\bf  p_{e^+}}$,
and we then use both 
${\bf \hat p_n}$ and ${\bf p_{e^+}}$ in reconstructing ${\bf \hat p_\nu}$.  See the Supplementary Material for more details on reconstructing ${\bf \hat p_\nu}$.

A key method for improving the angular resolution is timing. A typical neutron is captured within a few $\mu$s in the boron-loaded plastic scintillator. 
However, events where the neutron bounces multiple times between detector layers will be delayed, because as the neutron slows down, it takes time to cross the $1$ meter gap between layers. Lower timing cuts result in better angular resolution at the cost of a reduced rate.  A timing cut between the positron annihilation and neutron capture also helps discriminate from other random-coincidence backgrounds.
 
	Similarly, we require a minimum time delay between the positron event and the neutron capture equal to the amount of time required for the neutron to travel between layers.  
 This time delay depends on the reconstructed neutron momentum, but it is typically $\sim$$0.5$ $\mu$s for reactor-energy antineutrinos. 
 
	Another method for discriminating against events where the neutron has scattered significantly before capture is to require $ \cos \theta_{e n} = {\bf \hat p_{e^+}} \cdot {\bf \hat p_n}$ to be less than some minimum value, which we take to be zero in our analysis for definiteness.  This cut is more effective at antineutrino energies well above threshold; in the limit $E_{e^+} \gg 1.8$ MeV, the fraction of events with $ \cos \theta_{e n}>0$ shrinks to zero.  
	We only perform this cut when reconstructing ${\bf \hat p_\nu}$ from both ${\bf \hat p_n}$ and ${\bf p_{e^+}}$.
	See the supplementary material for details on the scattering kinematics and analysis.

As an illustration, we perform Monte Carlo simulations for $4$ MeV antineutrinos traveling in the direction ${\bf \hat p_\nu} = {\bf \hat z}$ (normal to the planes), in the notation of Fig.~\ref{Fig: 1}.    We generate $10^7$ IBD events in the target layer, for each target-layer thickness.
We define ${\bf p_\nu^\text{Rec.}} $ to be the reconstructed neutrino momentum vector and the angular error $\theta_\text{error}$ of the reconstruction by $\cos \theta_\text{error}=\bf \hat p_\nu \cdot \bf \hat p_\nu^\text{Rec.}$.

In Fig.~\ref{Fig: 3} we show the Monte-Carlo determined cumulative probability $P(\theta_\text{error} > \theta)$ that the angular error is greater than a value $\theta$.	
We reconstruct ${\bf p_\nu^\text{Rec.}} $ using the neutron's direction alone (dotted curves) and also by including the exact positron momentum (solid curves).
We illustrate the effect of a timing cut $\Delta t < 1$ $\mu$s (left panel) and $\Delta t < 6$ $\mu$s (right panel). The shorter timing cut results in better angular reconstruction, but less events are accepted. The inset plot on the right panel shows the fraction of events accepted as a function of the timing cut.

	The positrons are less likely to escape the target layer as the target-layer thickness is increased. For a $0.5$($1$)($2$) cm target, we find that $\sim$$45$\%($30$\%)($25$\%) of the positrons escape the target layer.

	The neutron-only reconstructions have similar errors across all target-layer thicknesses; these analyses are limited by the fact that we are neglecting the positron's momentum in reconstructing ${\bf p_\nu}$.
When we include the positron momenta in the reconstruction, the difference between target-layer thicknesses becomes clearer.
In Fig.~\ref{Fig: 3} it may be seen that thinner targets result in better angular resolution when including the positron's momentum in the analysis. For comparison, we also show the cumulative probability for a monolithic detector, consisting of the same boron-loaded plastic scintillator that is in the capture layers of the SANTA simulations.  In the monolithic simulations, we approximate ${\bf \hat p_\nu} \approx {\bf \hat p_n}$ using the exact neutron capture and IBD event locations. All of our SANTA target-layer thicknesses and ${\bf \hat p_\nu}$ reconstruction algorithms outperform the monolithic detector. See the Supplementary Material for examples with other antineutrino energies and incident angles.

{\it Discussion}---We have presented a novel design for a directional antineutrino detector that utilizes existing technology, and we have demonstrated its capability through Monte-Carlo simulations.
The detector works by segmenting the volume into alternating target and capture layers. The target layers are made thin enough for neutrons to escape with minimal elastic scattering. It is important to note, however, that non-directional IBD events may also be observed fully within the capture layers, making the detector dual purpose.
We have not attempted to optimize the parameters of the detector. There are a number of ways in which our example detector could be improved. The angular resolution increases with increasing distance between layers and decreasing target-layer thickness.

 The fact that there is empty space between detector layers does present a challenge for the scalability of the detector; a large-mass detector will necessarily take up a lot of physical space. However, the distance between layers and the thickness of the layers may be adjusted, depending on spacial constraints, required event rates, and desired angular resolution. 
 Liquid scintillator may also be used instead of plastic scintillator. 
 
 Depending on the application and detector size, it may be beneficial to include charged-particle tracking, such as a wire chamber, between layers.  This would help reduce backgrounds and measure ${\bf p_{e^+}}$.  Moreover, a $\sim$mT magnetic field  can be incorporated to differentiate charged particles by the curvature of their tracks within the gap. 
 
	It is also important to note that our detector has directional sensitivity to $\nu_e$-$e^-$ and $\bar \nu_e$-$e^-$ elastic scattering; we can reconstruct the momentum of the recoiling electron by tracking it through the detector.  We may extract directional information from elastic scattering events that take place in either the detector or the target layers. This makes our detector well suited, for example, for studying antineutrino-electron elastic scattering with an artificial antineutrino source, such as a nuclear reactor or IsoDAR~\cite{Conrad:2013sqa}.  Moreover, the elastic scattering events show up as double coincident signatures; the $e^-$ deposits energy in both layers, with a few ns delay.  With charged-particle tracking, the $e^-$ may also be tracked and identified between layers. These extra pieces of information help reduce background as compared to the same processes in monolithic detectors.

	A first-stage experiment might consist of a small detector placed near a nuclear reactor.  For example, consider a SANTA with a single \mbox{$2$ cm}-thick \mbox{$2$ m $\times$ $2$ m} target layer, between two \mbox{$6$ cm}-thick equal-area capture layers, placed $\sim$20 meters away from the core of a $3$ GW$_\text{th}$ nuclear reactor.
	Roughly $600$ IBD events would occur per day within the target layer, and an additional $\sim$$4 \times 10^3$ events would occur per day in the capture layers that could be used for non-directional detection.
	Such an experiment, while paving the way for larger detectors, would have immediate applications to nuclear reactor monitoring  
	and sterile neutrino searches.

	In a followup work, we will present a thorough detector simulation, including backgrounds and realistic detector properties, for a SANTA in the vicinity of a nuclear reactor.

{\it The authors would like to thank F. Calaprice, J. Conrad, J. Formaggio, P. Huber, S. Lee, M. Lisanti,  J. Spitz, M. Toups, C. Tully, and M. Vagins for helpful discussions. B.R.S. wishes to thank the Aspen Center for Physics, supported in part by the US NSF grant PHYS-1066293, and Princeton University for hospitality during this work. B.R.S was supported in part by a Pappalardo Fellowship in Physics at MIT and in part by the US Department of Energy under grant Contract Number DE-SC00012567. B.S. is supported by the US Department of Energy under grant Contract Number ER-41850.}

\vspace{-3.8in}

\vspace{4in}
\onecolumngrid
\vspace{0.3in}
\twocolumngrid
\def\bibsection{} 
\bibliographystyle{apsrev}
\bibliography{polarizedNeutrinos}

\newpage
\onecolumngrid
\begin{center}
\textbf{\large Supplementary Material}
\end{center}
\twocolumngrid
\setcounter{equation}{0}
\setcounter{figure}{0}
\setcounter{table}{0}
\setcounter{page}{1}
\makeatletter
\renewcommand{\theequation}{S\arabic{equation}}
\renewcommand{\thefigure}{S\arabic{figure}}

\begin{figure*}[htb]
	\leavevmode
	\begin{center}$
	\begin{array}{cc}
	\scalebox{.54}{\includegraphics{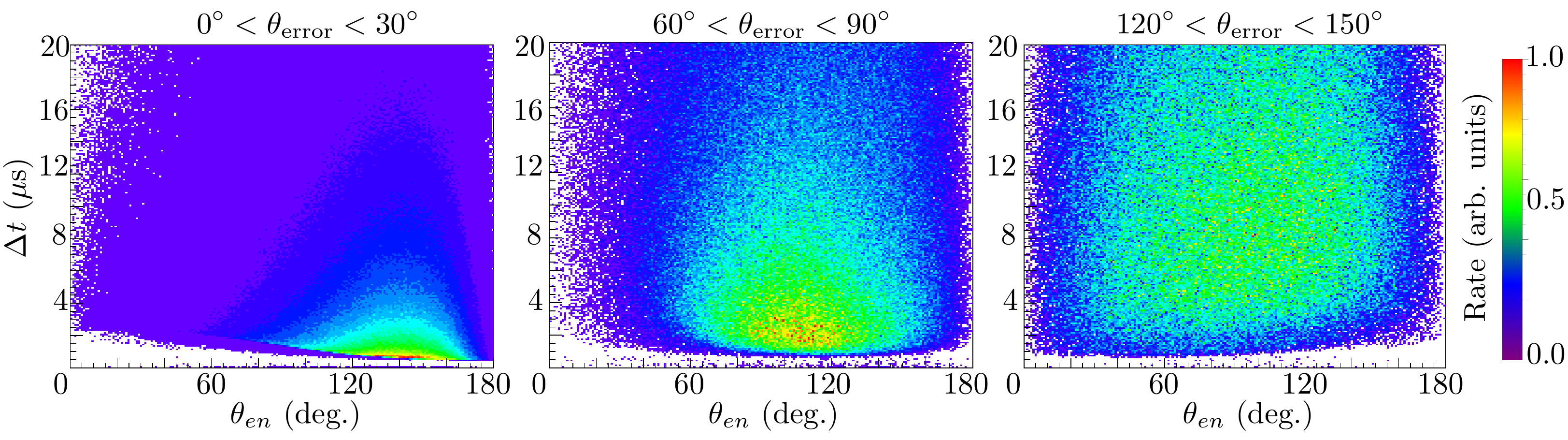}} \\
	 \end{array}$
	\end{center}
	\vspace{-.20cm}
	\caption{ Events with small $\theta_\text{error}$ tend to have small capture times $\Delta t$ and  $\theta_{en} > 90^\circ$. 
	 To illustrate this point, we generate $10^7$ IBD events in a SANTA with a 1 cm thick target layer.  The antineutrinos are incident normal to the target plane with $E_{\nu} = 4$ MeV.  We show $\Delta t$--$\theta_{en}$ histograms of the event rate for events with $\theta_\text{error}$ in three ranges: $0^{\circ}$--$30^{\circ}$, $60^{\circ}$--$90^{\circ}$, and $120^{\circ}$--$150^{\circ}$.  Note that each histogram is individually normalized.  
The histograms become more uniformly distributed as $\theta_\text{error}$ increases.
 By restricting $\Delta t$ and $\theta_{en}$, we achieve better accuracy.}
	\vspace{-0.15in}
	\label{Fig: Timing}
\end{figure*}  
 \begin{figure*}[tb]
\leavevmode
\begin{center}$
\begin{array}{cc}
\scalebox{.5}{\includegraphics{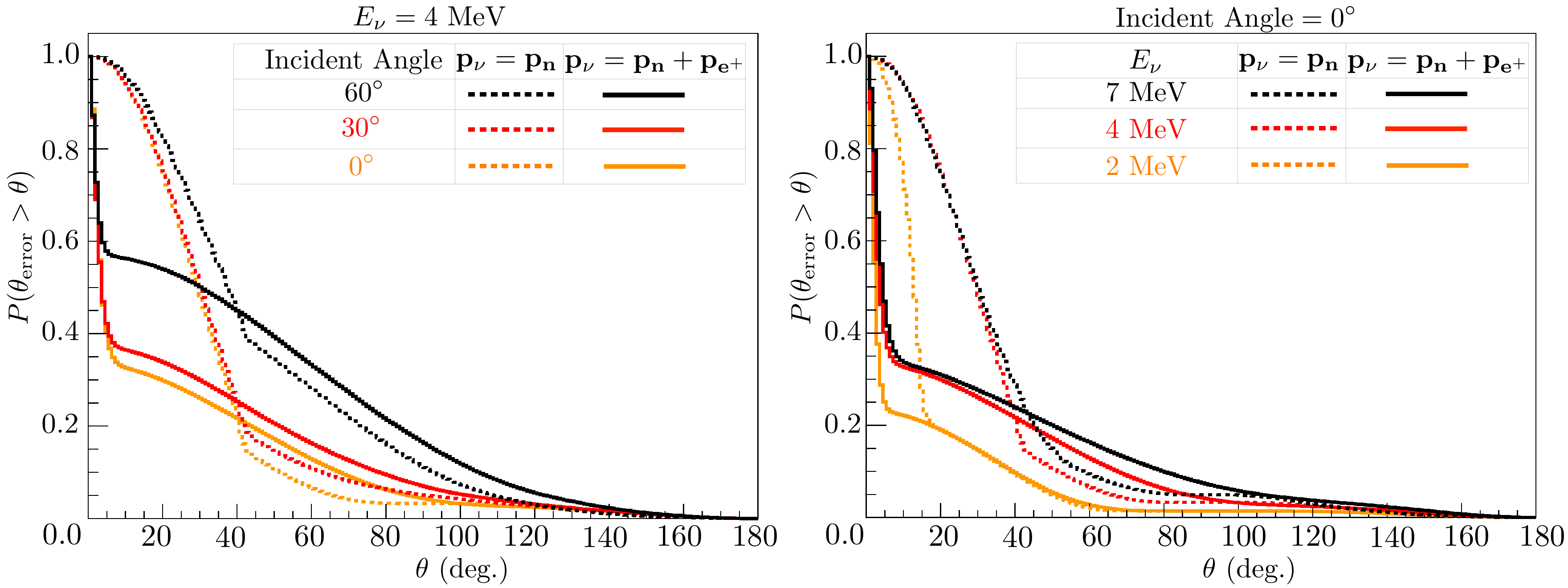}} \\
 \end{array}$
\end{center}
\vspace{-.50cm}
\caption{
We calculate the cumulative probability $P(\theta_\text{error} > \theta)$, with $\theta_\text{error}$ the angular error in reconstructing the momentum of the antineutrino, for different incident antineutrino angles (left panel) and energies (right panel).  We take a $1$ cm thick target, and we use the constraint $\Delta t < 2$ $\mu$s.  When including ${\bf p_{e^+}}$ in the reconstruction, we also require $\cos \theta_{en} < 0$.  The antineutrino direction is reconstructed best for low incident angles (normal to the plane) and low energies. 
  }
\vspace{-0.15in}
\label{Fig: Array}
\end{figure*}  

\section{Detector material}

We model our capture-layer scintillator material off the ELJEN EJ-254 scintillator~\cite{eljen} (5\% natural B by weight), which has hydrogen, carbon and $^{10}$B 
densities 
\es{}{
n_\text{H} &\approx 5.18 \times 10^{22} \, \, \text{cm}^{-3} \,, \, \,   n_\text{C} \approx 4.44 \times 10^{22}  \, \, \text{cm}^{-3} \,, \\
 n_{^{10}\text{B}} &\approx  5.68 \times 10^{20}  \, \, \text{cm}^{-3} \,.
}
For the target layer, we leave out the $^{10}$B, since it is not necessary.
These parameters are simply chosen to give a concrete example.

The desired thickness of the target layer may be estimated as follows.  The elastic-scattering cross-section for a $\sim$10 keV neutron on hydrogen(carbon) is \mbox{$\sigma_H \approx 20$}\mbox{($\sigma_C \approx 5$)} b.  Thus, the average distance a neutron will travel before its first elastic scattering is 
\es{}{
\ell_n = {1 \over n_H \sigma_H + n_C \sigma_C } \approx 0.8  \, \, \, \text{cm} \,.
}

\section{Reconstructing ${\bf p_\nu}$}

Given $E_{e^+}$, ${\bf \hat p_{e^+}}$, and ${\bf \hat p_n}$, we would like to reconstruct ${\bf p_\nu}$.  It is possible to work perturbatively in $E_{e^+} / m_n$, with $m_n$ the neutron's mass (see~\cite{PhysRevD.60.053003} for similar computations).
To leading order in $E_{e^+} / m_n$, we find that the neutron's kinetic energy may be written as
\es{NKE}{
K_n = &{E_\nu^2 \over 2 m_n} + {p_{e^+}^2 \over 2 m_n} \left[ \cos(2 \theta_{en})  \right. \\ &-\left.2 \cos \theta_{en} \sqrt{(E_\nu / p_{e^+})^2 -  \sin^2 \theta_{en} } \right] \,,
} 
with $p_{e^+}^2 = E_{e^+}^2 - m_e^2$ and $\cos \theta_{en} = {\bf \hat p_{e^+}} \cdot {\bf \hat p_n}$.  
The antineutrino's energy is $E_\nu \approx E_{e^+} + \Delta$, to leading order in $E_{e^+} / m_n$, with $\Delta \approx 1.29$ MeV the neutron-proton mass difference.
Then, by momentum conservation, 
\es{momC}{
{\bf p_\nu} = {\bf p_n} + {\bf p_{e^+}}  \,,
}
where ${\bf p_n} = \sqrt{2 m_n K_n} {\bf \hat p_n}$. 

It is important to note that $\cos \theta_{en}$ tends to be negative, as requiring $\cos \theta_{en}$ to be smaller than some minimum value is a method for improving the angular resolution of the detector. 
Using~\eqref{momC}, we may solve perturbatively for $\cos \theta_{en}$ in terms of $\cos \theta_{e \bar \nu} = {\bf \hat p_{e^+}} \cdot {\bf \hat p_\nu}$:
\es{cosen}{
\cos \theta_{e n} = - {p_{e^+} - E_\nu \cos \theta_{e \bar \nu} \over \sqrt{E_\nu^2 + p_{e^+}^2 - 2 E_\nu p_{e^+} \cos \theta_{e \bar \nu} }} \,.
}
This implies, for example, that $\cos \theta_{en}$ is negative so long as $\cos \theta_{e \bar \nu} < p_{e^+} / E_\nu$.

Heuristically, one may think of the IBD final-state positron as carrying away the antineutrino's energy, while the neutron carries away the momentum.  
In particular, the positron is emitted almost isotropically~\cite{PhysRevD.60.053003}: \mbox{$\langle \cos \theta_{e \bar \nu} \rangle \approx -0.034 p_{e^+} / E_{e^+}$}.  Combined with~\eqref{cosen}, for example, this implies that only a small subset of events will have positive $\cos \theta_{e n}$.  As $E_\nu$ becomes significantly large compared to the IBD threshold, the percentage of events with positive $\cos \theta_{e n}$ shrinks towards zero.  

The neutron, in contrast to the positron, tends to be emitted in the forward direction. By momentum conservation, the angle between the neutron and the antineutrino is necessarily smaller than an angle $\theta_M$, given by
\es{TM}{
\cos \theta_M = {\sqrt{2 E_\nu \Delta - (\Delta^2 - m_e^2) } \over E_\nu } \,,
}
with $m_e$ the mass of the electron.  The angle $\theta_M$ increases as $E_\nu$ increases from the threshold.  This means that directional information may 
be extracted from the neutron alone.  This is particularly relevant for IBD events near threshold, where $\theta_M$ is small and there is a higher chance that the positron will annihilate within the target layer.

\section{Timing and the neutron-positron angle}

The reconstructed neutron momenta may differ from the true neutron momenta because of neutron diffusion.  In our analysis, we use timing and $\theta_{en}$ constraints, with $\theta_{en}$ the angle between the positron and the reconstructed neutron,  to suppress events where the neutron has scattered significantly within the detector before capture.   

 Let $\Delta t$ denote the time between the IBD event and neutron capture 
 and  
  $\theta_\text{error}$ the angular error in the reconstruction of the antineutrino's momentum.   
To visualize the $\Delta t$ and $\theta_{e n}$ cuts used in the reconstruction, we construct $\Delta t$--$\theta_{en}$ histograms from Monte Carlo simulations in GEANT4.  We consider $4$ MeV antineutrinos incident normal to a $1$ cm thick target plane, and we generate $10^7$ IBD events within the target layer.  We consider three ranges for $\theta_\text{error}$: $0^{\circ}$--$30^{\circ}$, $60^{\circ}$--$90^{\circ}$, and $120^{\circ}$--$150^{\circ}$.  In Fig.~\ref{Fig: Timing} we show the corresponding $\Delta t$--$\theta_{en}$ histograms, where each histogram is constructed from the events with $\theta_\text{error}$ in the appropriate range.  Note that we reconstruct ${\bf p_\nu}$  using both the neutron and positron momenta.

 Most of the events in the $0^{\circ}$--$30^{\circ}$ histogram occur within $\sim$$4$ $\mu$s and with $\theta_{en} > 90^\circ$.
The $60^{\circ}$--$90^{\circ}$ and $120^{\circ}$--$150^{\circ}$ histograms, on the other hand, shows that the rate is more uniformly distributed in $\Delta t$ and $\theta_{en}$ when the angular error is larger.
 Thus, by imposing constraints on $\cos \theta_{en}$ and $\Delta t$, the reconstruction accuracy may be improved.

\section{Varying the antineutrino energy and incident angle}

It is useful to understand how the detector performs for different antineutrino energies and incident angles.
As the incident angle with respect to the normal direction of the target plane is increased, the resulting neutrons will tend to traverse longer distances in the target layer.  Moreover, neutrons that leave the target at a large angle take more time to reach the capture layer and are more likely to be reflected out of the capture layer.
As a result, the detector performs best when the antineutrinos are incident normal to the plane. 

Similarly, as the energy of the antineutrino increases, so does the average angle between the neutron and the antineutrino.  
The detector therefore performs best at low antineutrino energies, near the IBD threshold. 
 
To validate the intuition above, we perform a variety of detector Monte Carlo simulations, with the results shown in Fig.~\ref{Fig: Array}.  In each of these simulations, we consider a SANTA with a 1 cm thick target layer. 
We generate $10^7$ IBD events in the target layer in each case, and we impose the constraint $\Delta t < 2$ $\mu$s and, where applicable, $\cos \theta_{en} < 0$.  

 In the left panel of Fig.~\ref{Fig: Array}, we consider $4$ MeV antineutrinos with incident angles $0^{\circ}$, $30^{\circ}$, and $60^{\circ}$.  We reconstruct the antineutrino's momentum using the neutron's momentum alone and also by including the positron's momentum. The results show that the angular resolution is best when the antineutrinos are incident normal to the plane.
In the right panel we consider antineutrinos incident normal to the plane with energies $2$, $4$, and $7$ MeV; as expected, the angular resolution increases with decreasing antineutrino energy.

\end{document}